# *Driven Time Crystal in Low-Symmetry ENZ Conductors*

*Mário G. Silveirinha*[(1)*]

[(1)] *University of Lisbon–Instituto Superior Técnico and Instituto de Telecomunicações, Avenida Rovisco Pais, 1, 1049-001 Lisboa, Portugal*

**Abstract**

In recent years, epsilon-near-zero (ENZ) materials have attracted a great deal of attention in nonlinear optics, as they combine field enhancement with strong, ultrafast nonlinearities. In particular, transparent conducting oxides, such as ITO, have emerged as a promising class of materials, and have been extensively exploited to achieve temporal optical responses varying on the femtosecond scale using an optical pump. Most of the solutions discussed so far in the literature rely on effective $\chi^{(3)}$ modulations, wherein the dominant material response is controlled by the envelope of the optical pump. Here, it is shown that low-symmetry conductors can provide an interesting alternative to transparent conducting oxides and a more natural implementation of time-crystalline behavior in optical systems with optical-cycle modulation. I demonstrate that ENZ confinement, combined with the strong anomalous-velocity nonlinearity of low-symmetry conductors, enables a subwavelength nanoparticle to develop a time-crystalline response under optical pumping. For sufficiently strong pumping, this response can overcome dissipative losses and lead to parametric amplification. Furthermore, the pump can strongly tailor the scattering and extinction of a weak probe, and in extreme cases render the extinction negative. In this regime, the driven nanoparticle effectively amplifies the probe beam.

[*] To whom correspondence should be addressed: E-mail: mario.silveirinha@tecnico.ulisboa.pt

# I. Introduction

Standard dielectric optical materials exhibit relatively weak nonlinearities [1]. Therefore, typical nonlinear optical applications, such as parametric amplification or harmonic generation, usually require intense pump fields that can induce a sufficiently large nonlinear polarization. This inherently weak nonlinear response is a consequence of the fact that the optical response of these materials is governed by bound electrons confined to a region of space with atomic dimensions. Due to the strong Coulomb fields that bind the electrons to the atoms, the effective displacement of the electrons during one optical cycle is tiny, and their sensitivity to anharmonic effects is small [1]. Consequently, a strong applied optical field typically affects the material response only perturbatively. In fact, a strategy to enable strong nonlinearities is to confine carriers in shallow asymmetric electronic potentials, and this is typically achieved through band engineering in semiconductor quantum wells [2-4].

Due to the above-mentioned limitations, alternatives to ordinary insulating dielectrics have been considered. In particular, an interesting paradigm based on nonlinear conducting systems operated near the zero crossing of the real part of the linear permittivity, known as the epsilon-near-zero (ENZ) regime [5-6], has emerged during the last decade [7-9]. It has been shown that transparent conducting oxides (TCOs), such as indium tin oxide (ITO), can enable ultrafast and strong modulations of the refractive index at optical frequencies [10-13]. These properties may be attributed to a combination of factors [14].

First, Bloch electrons behave approximately as free particles in a conductor. A useful picture is that they move in a shallow and nonuniform periodic potential created by the ionic lattice. Consequently, in an optical cycle, they can swing over much larger distances than in a dielectric. This enables them to probe, for example, non-parabolic effects in the electronic band structure, which describe deviations from the free-electron model and correlate with an energy-dependent effective mass [15]. This specific process has an electronic origin, and the corresponding modulation can be ultrafast [16]. Slower dynamics are also observed in experiments and are typically attributed to thermalization effects and electron–phonon relaxation [16-18].

The second factor originates from the fact that in the ENZ regime, the electric field can be enhanced inside the material relative to the applied field. This means that the same intensity of the applied field may produce a stronger nonlinear response when $\varepsilon \approx 0$ [14]. Physically, this field enhancement can be attributed in part to the fact that, in the ENZ regime, the counter-propagating (bulk) eigenwaves coalesce at $k$=0 [19]. As in other passive waveguiding systems with exceptional points [20], this coalescence may correlate with resonant behavior, often a zero-order resonance [21]. The distinctiveness of ENZ materials in this regard is that the corresponding resonant behavior may be tolerant to geometry changes [5-6], due to its quasi-static nature.

Finally, the third reason for the relevance of nonlinear conducting materials is that the sensitivity of optical fields to variations in the material permittivity tends to be especially strong when the permittivity is near zero [14]. Indeed, when $\varepsilon$ crosses zero, the relative change in the refractive index can be large and thus have a stronger impact on scattering processes and absorption than in other spectral regions.

In recent years, the pursuit of time-varying optics, in which the material response is required to vary on a timescale comparable with the optical cycle [22-24], has kindled further interest in TCOs. Indeed, some of the most impressive consequences of abrupt material-response switching, including time refraction, double-slit time diffraction and signatures of time reflection, have been experimentally demonstrated using this class of materials in the optical domain [25-28].

Time-varying platforms break continuous time-translation symmetry, creating the possibility of injecting energy into the system through the modulation mechanism [22]. When the modulation is rooted in an optical nonlinearity, this phenomenon is closely related to conventional optical parametric amplification [24, 29]. In simple terms, in the nondispersive scenario, whenever the permittivity undergoes a downward transition, the modulation injects energy into the system, whereas when the permittivity experiences an upward transition, the modulation extracts energy [30]. In both processes, the energy exchange is proportional to the instantaneous electric energy density in the material. Thus, when the upward and downward transitions alternate in time, defining time-crystalline behavior, the process that dominates depends on the relative phase between the oscillations of the electric energy density and the modulation. In particular, significant amplification of an optical signal requires, in principle, that the modulation cycle be repeated multiple times.

Typical TCOs used in time-varying photonics are centrosymmetric materials, and thus their dominant nonlinear response is described by a third-order nonlinear susceptibility $\chi^{(3)}$. Most of the experimental work reported so far modulates the optical pump intensity to create an abrupt temporal interface [25-28]. Unfortunately, intensity modulation typically involves significant repopulation of the electronic states, accompanied by slow relaxation processes when the pump is switched off, which prevent the modulation cycle from being repeated on a timescale short enough to achieve parametric amplification [16-18]. Thus, it does not appear feasible to engineer a photonic time-crystalline phase using a periodic modulation of the pump envelope in typical TCO platforms in the optical domain.

As discussed in recent works [31, 32, 33], low-symmetry materials, specifically systems with broken spatial inversion symmetry, may offer a more sensible route to engineer a time-crystalline phase in the infrared and optical domains. Such materials may possess a second-order nonlinear susceptibility $\chi^{(2)}$, and thereby naturally allow for optical-cycle modulation with a constant-envelope pump [31, 32, 33]. Because the $\chi^{(2)}$ response is the lowest-order nonlinearity, and thus generally the strongest, these systems are arguably the natural platforms for this purpose [31].

I note in passing that $\chi^{(3)}$ systems may also provide optical-cycle modulation when operated with a constant-envelope pump [30, 34]. However, time crystals formed in these systems have a temporal modulation frequency twice the pump frequency ($\Omega = 2\omega_{\text{pump}}$). Thus, the spectral range where momentum gaps are expected to emerge, centered near $\omega \approx \Omega / 2$, overlaps the pump frequency, complicating the spectral separation of the pump and probe and limiting the range of possible applications. In contrast, for a $\chi^{(2)}$ modulation, the time-crystal frequency is $\Omega = \omega_{\text{pump}}$, so that the pump does not overlap the most relevant spectral range.

Motivated by these general considerations, in this article I investigate the temporal modulation of low-symmetry conductors operated close to the ENZ point. This class of materials has become a focus of attention in the condensed-matter community after the proposal and experimental verification of the nonlinear Hall effect [35-37]. Such systems, which include certain Weyl semimetals and transition-metal dichalcogenides, behave as ordinary conductors in the linear regime. Due to their broken microscopic inversion symmetry, the transport of Bloch electrons in these systems is influenced by an anomalous velocity contribution controlled by the quantum geometry of the material [38-39]. It is this anomalous velocity mechanism that underpins the nonlinear Hall effect. Specifically, the electric current density includes a second-order nonlinear contribution that is proportional to the applied electric field and to a material dependent tensor known as the Berry curvature dipole (BCD).

Second-order nonlinearities in conductors have mostly been discussed in the literature in the context of electron transport at very low frequencies [35], and more recently in the context of second harmonic generation [40-42]. In particular, a few experimental studies have demonstrated that the second-order nonlinearity in these materials can significantly exceed that of standard nonlinear optical materials such as GaAs [40, 41]. Similar to TCOs, a large intraband nonlinearity in low-symmetry conductors can be attributed to the fact that Bloch electrons are weakly confined by the shallow potential of the ionic lattice. Unlike conventional TCOs, however, the ionic potential lacks inversion symmetry, thereby enabling a second-order nonlinear response.

Previous work more closely related to nanophotonics has focused primarily on electro-optic effects [43-51]. In particular, it has been shown that Berry-dipole materials under a DC electric bias can provide nonreciprocal and non-Hermitian responses that may feature optical gain [43-48, 51]. Remarkably, in these systems, the gain is controlled by the optical-field polarization. For some eigenpolarizations determined by the static electric bias and crystal symmetry, the electro-optic contribution to the optical response may result in increased dissipation, whereas for other eigenpolarizations, it may result in reduced dissipation or even gain under a sufficiently strong bias [46].

Given the strong nonlinearities that characterize these conducting systems and the fact that they can also exhibit a plasmonic-type response, it is natural to wonder whether they can be useful for ENZ nonlinear nanophotonics [33]. A recent study for extended low-symmetry conductors showed that a traveling optical pump may generate a spacetime modulation and collective parametric resonances near the ENZ point [32]. However, in an extended medium the pump itself propagates through the dissipative material and is therefore subject to attenuation.

This consideration motivates configurations in which the nonlinear interaction is localized in space, avoiding the need for the pump to propagate over an extended distance through the lossy material. Here, I investigate such a problem and focus on the time-varying response of a finite subwavelength resonator and on its parametric and scattering properties.

The article is organized as follows. In Sect. II, I introduce the BCD-Drude model, a time-domain transport model for the second-order nonlinear response of low-symmetry conductors. I demonstrate that this model predicts exactly the same electrodynamics as the standard frequency-domain nonlinear-optics formalism, within the semiclassical approximation. In Sect. III, I study the electromagnetic response of a subwavelength BCD nanoparticle in a pump-probe configuration. Using a quasistatic approximation, I derive a system of master equations that determines the time evolution of the particle electric dipole moment in terms of the pump and probe fields. In particular, through the linearization of the master equations, I demonstrate that a strong pump can induce a driven time-crystalline response. Then, in Sect. IV, I apply the developed framework to investigate parametric amplification and the effect of the pump field on the scattering response of the nanoparticle. I also discuss conditions under which the extinction can become negative. Finally, Sect. V presents a brief summary of the article and discusses future perspectives.

## II. Nonlinear Model

In this section, I introduce the BCD-Drude model and discuss its range of applicability to low-symmetry conductors.

### *A. Time-domain BCD-Drude model*

The second-order nonlinear response of a material is usually modeled in the frequency domain by specifying the response associated with different frequency channels. The archetypal relation between the second-order current density and two generic field harmonics with frequencies $\omega_1$ and $\omega_2$ is

$$\mathbf{j}^{(2)}_{\omega_1+\omega_2,\alpha} = \sigma^{(2)}_{\alpha\beta\nu}\left(-\omega_3;\omega_1,\omega_2\right) E_{\omega_1,\beta} E_{\omega_2,\nu}, \tag{1}$$

where $\sigma^{(2)}_{\alpha\beta\nu}\left(-\omega_3;\omega_1,\omega_2\right)$ is the second-order conductivity tensor and $\omega_3 = \omega_1 + \omega_2$. Previous works have used microscopic-response-theory methods to express the second-order conductivity $\sigma^{(2)}_{\alpha\beta\nu}$ directly in terms of the electronic band structure of the material [52-55].

When the relevant fields have oscillation frequencies well below the electronic interband transition frequencies, $|\omega_1| << \Delta/\hbar$, $|\omega_2| << \Delta/\hbar$ and $|\omega_1+\omega_2| << \Delta/\hbar$, with $\Delta$ the relevant gap energy and $\hbar$ the reduced Planck constant, the leading second-order response can be expressed using a single material-specific tensor $\overline{\mathbf{D}}$ [35, 55-56], known as the Berry-curvature dipole [35]. This is the semiclassical regime considered in this article. In this discussion, it is implicit that the material is described by a time-reversal-symmetric Hamiltonian. When time-reversal symmetry is broken, contributions beyond the Berry dipole can arise [56]. Furthermore, it has been shown that, for some systems, the second-order current may also include contributions involving both the electric and magnetic fields (magnetoelectric response) [49-51], but I neglect these here.

Interestingly, within the discussed approximations, the nonlinear dynamics of the material can be formulated directly in the time domain through a BCD-Drude model. The model generalizes the standard Drude formalism.

Similar to the Drude model, the Bloch electrons near the Fermi level are characterized by an effective quasi-momentum $\mathbf{p}$ that responds to an applied electric field as:

$$\frac{\partial \mathbf{p}}{\partial t} + \Gamma \mathbf{p} = -e\mathbf{E}. \qquad (2)$$

Here, $\Gamma$ denotes the collision frequency and $e > 0$ the elementary charge. The electric current density $\mathbf{j} = -en\mathbf{v}$ is governed by the electron density $n$ and by the velocity $\mathbf{v}$. For standard centrosymmetric conducting materials, the velocity and the quasi-momentum can be linked as $\mathbf{v} = \mathbf{p}/m^*$, with $m^*$ the effective mass determined by the curvature of the electronic energy dispersion. For simplicity, I consider an isotropic (scalar) effective mass.

The important point is that in a low-symmetry conductor the velocity of the Bloch electrons may include an additional contribution, known as the anomalous velocity [38-39]. The anomalous velocity is governed by the quantum geometry of the material through the Berry curvature [39]. In the context of electron transport, the relevant contribution is determined by a weighted average of the Berry curvature of electronic states near the Fermi level. Specifically, the effective anomalous velocity can be expressed as $\mathbf{v}_{\text{an}} = -\frac{e}{n\hbar^2}\left(\mathbf{p}\cdot\overline{\mathbf{D}}\right)\times\mathbf{E}$, where $\overline{\mathbf{D}}$ is the Berry-curvature dipole. This tensor effectively describes the "dipole moment" of each scalar component of the Berry curvature near the Fermi surface (see Fig. 1, leftmost panel). For a three-dimensional material, the Berry curvature dipole is dimensionless and its specific structure is material-dependent [46, 57]. It can be nontrivial only for conducting materials that break inversion symmetry.

*Quantum geometry* *Time-varying photonics*

**Fig. 1 Left:** Illustration of the origin of the Berry curvature dipole. The figure shows a density plot of the *y*-component of the Berry curvature, $\Omega_{\mathrm{B},y}$, on a plane in momentum space near the Fermi surface of a low-symmetry conductor. **Middle:** The Berry curvature dipole gives rise to an anomalous-velocity contribution responsible for the nonlinear current. **Right:** Under optical pumping, this nonlinearity generates a time-dependent effective conductivity described by the BCD-Drude model. The nonlinearity enables frequency mixing between the pump and a weak probe signal, generating Floquet sidebands. A time-periodic pump induces a driven time-crystalline response in a subwavelength spheroidal nanoparticle.

The total electric current density in the material can thus be expressed as:

$$\mathbf{j} = -ne\frac{\mathbf{p}}{m^*} + \frac{e^2}{\hbar^2}\left(\mathbf{p}\cdot\overline{\mathbf{D}}\right)\times\mathbf{E}\,. \tag{3}$$

The second term is the anomalous velocity contribution. It is clearly nonlinear as it involves products of components of the electric field and of the quasi-momentum.

Equations (2)-(3) define a time-domain model for electron transport in low-symmetry conductors. This framework, which I will refer to as the BCD-Drude model, was first introduced phenomenologically in Ref. [58] to describe electro-optic effects [43, 44]. However, as discussed in further detail in Appendix A, it actually provides an exact account of the BCD nonlinear response of these materials, within the semiclassical approximation. Indeed, it aggregates in a compact manner all the second-order nonlinear frequency-mixing processes, including second-harmonic generation, electro-optic effects, and parametric driving. The model is valid provided the largest relevant field frequency and the relaxation rate $\Gamma$ are well below the interband transition frequency $\Delta/\hbar$.

To conclude this subsection, I note that the second-order current associated with the Berry curvature dipole does not contribute to the work done by the electric field, because $\mathbf{j}_{\mathrm{BCD}} \cdot \mathbf{E} = 0$ with $\mathbf{j}_{\mathrm{BCD}} = \frac{e^2}{\hbar^2}\left(\mathbf{p}\cdot\overline{\mathbf{D}}\right)\times\mathbf{E}$. Thus, the Poynting theorem for these systems does not explicitly involve the Berry-curvature dipole [58].

### *B. Some material considerations*

There is a compromise between the strength of the Berry-curvature dipole and the spectral range over which the semiclassical approximation is applicable. As shown in Appendix B, the magnitude of $\overline{\mathbf{D}}$ is bounded by a quantity that scales as $1/\Delta^2$. Here, $\Delta$ is the minimum energy separation between the partially occupied band responsible for the electronic transport and the neighboring bands. Thus, a small interband separation generally favors a large $\overline{\mathbf{D}}$. On the other hand, the same energy scale sets the range of validity of the low-frequency approximation underlying the BCD-Drude model [Eqs. (2)-(3)], which, as discussed in the previous subsection, requires that $|\omega| << \Delta/\hbar$.

This observation provides useful material guidelines. Low-symmetry doped insulators and semiconductors may be particularly interesting, because doping creates the Fermi surface required for the BCD response while the underlying band structure can retain a relatively large interband separation. In particular, *n*-type doping partially fills the conduction band and can give rise to a BCD for certain crystal point groups. Trigonal tellurium is a representative example of this class of materials [44].

It should be noted that the interband transition energy scale is not a hard upper limit on exploiting the nonlinear response of these materials. Even though outside the semiclassical regime additional nonlinear mechanisms become important, the full second-order response may remain exceptionally large and can even be enhanced by resonant electronic transitions. This behavior is well illustrated by TaAs [40, 41] and BiTeI [59]. A related mechanism can also enhance the nonlinear response of intrinsic semiconductors [60]. However, it is important to keep in mind that above the threshold for interband transitions the response of the material may become prohibitively lossy.

In the present article, I restrict the analysis to the spectral range below the relevant one-photon interband transition scale, where the semiclassical model provides the leading-order description of the material response. This spectral range also includes the regime where a plasmonic response emerges.

Bismuth telluroiodide (BiTeI) provides a reasonable compromise between a sizable Berry dipole, operation below the interband transition frequency, and moderate plasmonic loss in the infrared range [32, 61, 62]. This polar semiconductor has a valence-conduction gap of approximately 0.36 eV, corresponding to a frequency scale of about 90 THz. At the same time, experiments on bulk BiTeI report a collision frequency on the order of $\Gamma/(2\pi) \approx 3.5\,\text{THz}$ at low temperatures [62]. The ENZ crossing frequency $\omega_{\text{ENZ}}/(2\pi)$ can in principle be tuned in a relatively broad range of frequencies in the mid-infrared, roughly 10 to 40THz, while retaining a significant Berry curvature dipole [61, 62]. Thus, the ENZ frequency is naturally in the mid-infrared range and can be chosen well below the relevant interband threshold.

In the most interesting configuration for time-modulation, the pump frequency ($\Omega$) is about twice the ENZ frequency, $\Omega \approx 2\omega_{\text{ENZ}}$ (see Sect. IV). Depending on $\omega_{\text{ENZ}}$, the pump frequency may approach the valence-conduction band interband threshold. The proximity to the interband threshold may imply some deviations from the BCD-Drude model, but the corresponding corrections may actually be beneficial, as the full second-order nonlinear response may be enhanced as an interband resonance is approached from below [54, 63].

## III. Master Equation and Driven Time-Crystal Response

In the following, I develop a quasistatic model for the electromagnetic response of a homogeneous subwavelength spheroidal BCD nanoparticle in a pump-probe configuration, as illustrated in the rightmost panel of Fig. 1. The nanoparticle is illuminated by a strong pump field. I begin by characterizing its linear response to the optical pump.

### *A. Linear response to the optical pump*

In the linear regime, I ignore the second-order current associated with the Berry curvature dipole. Then, the material is simply described by a Drude dispersive model:

$$\varepsilon(\omega)=\varepsilon_\infty\left(1-\frac{\omega_{\mathrm{ENZ}}^2}{\omega(\omega+i\Gamma)}\right). \tag{4}$$

Here, $\omega_{\mathrm{ENZ}}=\sqrt{\frac{e^2 n}{m^* \varepsilon_0 \varepsilon_\infty}}$ is the screened plasma frequency (ENZ point) and $\varepsilon_\infty$ is the relative permittivity of the dielectric background, which includes the contribution from bound electrons. For BiTeI, one can use $\varepsilon_\infty \sim 13$ [62]. The actual response of BiTeI is anisotropic, but for conceptual clarity I ignore the anisotropy here.

As is well known, a subwavelength spheroidal particle behaves as an electric dipole. For a prolate spheroidal geometry both the interior electric field $\mathbf{E}^{\mathrm{int}}$ and the polarization vector $\mathbf{P}^{\mathrm{int}}$ can be assumed spatially uniform inside the particle [64]. Furthermore, the interior field is related to the incident field $\mathbf{E}^{\text{inc}}$ and polarization vector $\mathbf{P}^{\text{int}}$ as:

$$\mathbf{E}^{\text{int}} = \mathbf{E}^{\text{inc}} - \mathbf{L} \cdot \frac{\mathbf{P}^{\text{int}}}{\varepsilon_0}, \tag{5}$$

where $\mathbf{L}$ is a depolarization tensor that depends on the geometry of the particle and determines the depolarization field [64]. This result is very standard, but the point that I wish to emphasize here is that it is still applicable in a scenario where the nanoparticle response becomes nonlinear, provided the relation between $\mathbf{P}^{\text{int}}$ and $\mathbf{E}^{\text{int}}$ remains local in space and the particle is homogeneous. This property will be used in the next subsection.

For the spheroidal particle sketched in Fig. 2a with symmetry axis along *z*, the tensor $\mathbf{L}$ has a uniaxial structure [64]:

$$\mathbf{L} = L_t \left( \hat{\mathbf{x}} \otimes \hat{\mathbf{x}} + \hat{\mathbf{y}} \otimes \hat{\mathbf{y}} \right) + L_{zz} \hat{\mathbf{z}} \otimes \hat{\mathbf{z}}, \qquad \text{with} \quad 2L_t + L_{zz} = 1. \tag{6}$$

In particular, when the two particle semiaxes, $R_t$ and $R_z$, are identical (see Fig. 2a), the particle becomes spherical and $L_t = L_{zz} = 1/3$. Another relevant case corresponds to an elongated spheroid with $R_t << R_z$ (cigar- or needle-shaped) for which $L_t \approx 1/2$ and $L_{zz} \approx 0$ [64].

Evidently, in the linear regime the interior electric field and polarization vectors are linked in the frequency domain as $\mathbf{P}_\omega^{\text{int}} = \varepsilon_0 \left( \varepsilon(\omega) - 1 \right) \mathbf{E}_\omega^{\text{int}}$. Using this result, one can readily find the response to an incident optical pump described by the field:

$$\mathbf{E}_{\text{pu}} = \frac{1}{2} \mathbf{E}_{\text{pu}0} e^{-i\Omega t} + \frac{1}{2} \mathbf{E}_{\text{pu}0}^* e^{+i\Omega t}. \tag{7}$$

The corresponding induced interior field is:

$$\mathbf{E}_{\text{pu}}^{\text{int}} = \frac{1}{2} \mathbf{E}_{\text{pu}0}^{\text{int}} e^{-i\Omega t} + \frac{1}{2} \mathbf{E}_{\text{pu}0}^{\text{int}*} e^{+i\Omega t}, \qquad \text{with} \quad \mathbf{E}_{\text{pu}0}^{\text{int}} = \left( \mathbf{1} + \left( \varepsilon(\Omega) - 1 \right) \mathbf{L} \right)^{-1} \cdot \mathbf{E}_{\text{pu}0}. \tag{8}$$

Here, $\varepsilon(\Omega)$ is the Drude relative permittivity [Eq. (4)] evaluated at the pump frequency $\Omega$. For future reference, I note that the pump-induced quasi-momentum of the carriers [see Eq. (2)] is given by $\mathbf{p}_{\rm pu} = \frac{-e}{\Gamma}\tilde{\mathbf{E}}_{\rm pu}$, where

$$\tilde{\mathbf{E}}_{\rm pu} = \frac{1}{2}\tilde{\mathbf{E}}_{\rm pu0}e^{-i\Omega t} + \frac{1}{2}\tilde{\mathbf{E}}^{*}_{\rm pu0}e^{+i\Omega t}, \qquad \text{with } \tilde{\mathbf{E}}_{\rm pu0} = \frac{\Gamma}{\Gamma - i\Omega}\mathbf{E}^{\rm int}_{\rm pu0}. \tag{9}$$

### *B. Master system of equations*

I now turn to the general nonlinear response of the nanoparticle. As already mentioned in the previous subsection, Eq. (5) remains valid in the nonlinear regime. As shown next, together with the transport equations (2)-(3), it describes the complete time-domain dynamics of the subwavelength particle.

To this end, it is convenient to introduce a polarization vector $\mathbf{P}_{\rm c}^{\rm int}$ associated with the free carriers in the nanoparticle, such that $\mathbf{j} = \partial_t \mathbf{P}_{\rm c}^{\rm int}$. Then, the total polarization vector in the nanoparticle is:

$$\mathbf{P}^{\rm int} = \varepsilon_0\left(\varepsilon_\infty - 1\right)\mathbf{E}^{\rm int} + \mathbf{P}_{\rm c}^{\rm int}. \tag{10}$$

The first term describes the contribution of the bound electrons, whereas the second describes that of the free carriers. Substituting this result into Eq. (5), one obtains an explicit formula for the interior field:

$$\mathbf{E}^{\rm int} = \left(\mathbf{1} + \left(\varepsilon_\infty - 1\right)\mathbf{L}\right)^{-1}\cdot\left(\mathbf{E}^{\rm inc} - \mathbf{L}\cdot\frac{\mathbf{P}_{\rm c}^{\rm int}}{\varepsilon_0}\right). \tag{11a}$$

A differential equation in time for $\mathbf{P}_{\rm c}^{\rm int}$ can now be obtained by replacing $\mathbf{j} = \partial_t \mathbf{P}_{\rm c}^{\rm int}$ into the transport equations (2)-(3):

$$\frac{d}{dt}\frac{\mathbf{P}_{\mathrm{c}}^{\mathrm{int}}}{\varepsilon_0} = \frac{\varepsilon_\infty \omega_{\mathrm{ENZ}}^2}{\Gamma}\tilde{\mathbf{E}} + \frac{-e^3}{\varepsilon_0 \hbar^2 \Gamma}\left(\tilde{\mathbf{E}}\cdot\overline{\mathbf{D}}\right)\times\mathbf{E}^{\mathrm{int}} . \quad (11b)$$

$$\frac{d\tilde{\mathbf{E}}}{dt} + \Gamma\tilde{\mathbf{E}} = \Gamma\mathbf{E}^{\mathrm{int}} . \quad (11c)$$

where for convenience I introduced $\tilde{\mathbf{E}} = \frac{\Gamma}{-e}\mathbf{p}$ which has units of electric field.

The master system of equations (11) describes exactly the nonlinear dynamics of the subwavelength particle within the BCD-Drude model and the quasistatic approximation. It defines a first-order system of nonlinear differential equations with respect to the unknowns $\mathbf{P}_{\mathrm{c}}^{\mathrm{int}}$ and auxiliary field $\tilde{\mathbf{E}}$. Note that $\tilde{\mathbf{E}}$ is proportional to the effective quasi-momentum of the carriers. The vectors $\mathbf{P}_{\mathrm{c}}^{\mathrm{int}}$ and $\tilde{\mathbf{E}}$ respond to the incident field $\mathbf{E}^{\mathrm{inc}}$.

Furthermore, combining Eqs. (5) and (10) and solving for $\mathbf{P}^{\mathrm{int}}$ one can write an explicit formula for the induced electric dipole moment in terms of the incident field and $\mathbf{P}_{\mathrm{c}}^{\mathrm{int}}$:

$$\frac{\mathbf{p}_{\mathrm{dip}}}{\varepsilon_0} = V\left(\mathbf{1} + (\varepsilon_\infty - 1)\mathbf{L}\right)^{-1}\cdot\left[(\varepsilon_\infty - 1)\mathbf{E}^{\mathrm{inc}} + \frac{\mathbf{P}_{\mathrm{c}}^{\mathrm{int}}}{\varepsilon_0}\right], \quad (12)$$

where $V$ is the volume of the particle.

BiTeI has a crystal symmetry axis, which I will assume to be oriented along the $z$-direction. Then, the corresponding BCD tensor takes the form $\overline{\mathbf{D}} = D_{\mathrm{B}}\left(\hat{\mathbf{x}}\otimes\hat{\mathbf{y}} - \hat{\mathbf{y}}\otimes\hat{\mathbf{x}}\right)$, with $D_{\mathrm{B}} \sim 0.01$ [61]. As the tensor is antisymmetric, it can be expressed in terms of a cross-product: $\overline{\mathbf{D}} = -D_{\mathrm{B}}\hat{\mathbf{z}}\times\mathbf{1}$. In particular, the term $\tilde{\mathbf{E}}\cdot\overline{\mathbf{D}} = \overline{\mathbf{D}}^{\mathrm{T}}\cdot\tilde{\mathbf{E}} = -\overline{\mathbf{D}}\cdot\tilde{\mathbf{E}}$ in Eq. (11b) can be expressed as $\tilde{\mathbf{E}}\cdot\overline{\mathbf{D}} = D_{\mathrm{B}}\hat{\mathbf{z}}\times\tilde{\mathbf{E}}$.

### *C. Driven time-crystal response*

Next, I consider that a strong optical pump illuminates the nanoparticle. Its electric field $\mathbf{E}_{\mathrm{pu}}(t)$ is given by Eq. (7). Suppose that this system is probed by a much weaker time-harmonic field $\mathbf{E}_{\mathrm{probe}}(t)$ with a frequency $\omega$ different from the pump (Fig. 1, rightmost panel). Then, to leading order, the response to the probe can be found through linearization of the master system of Eqs. (11). Specifically, write $\mathbf{E}^{\mathrm{inc}} = \mathbf{E}_{\mathrm{probe}} + \mathbf{E}_{\mathrm{pu}}$, $\tilde{\mathbf{E}} = \tilde{\mathbf{E}}_{\mathrm{probe}} + \tilde{\mathbf{E}}_{\mathrm{pu}}$ and $\mathbf{P}_{\mathrm{c}}^{\mathrm{int}} = \mathbf{P}_{\mathrm{c,probe}}^{\mathrm{int}} + \mathbf{P}_{\mathrm{c,pu}}^{\mathrm{int}}$, where the response fields with the "pu" index are exact solutions obtained without the probe illumination. Substituting these decompositions into system (11) and retaining only terms that are linear in the probe fields, one finds:

$$\frac{d}{dt}\frac{\mathbf{P}_{\mathrm{c,probe}}^{\mathrm{int}}}{\varepsilon_0} = \frac{\varepsilon_\infty \omega_{\mathrm{ENZ}}^2}{\Gamma}\tilde{\mathbf{E}}_{\mathrm{probe}} - \frac{e^3}{\varepsilon_0 \hbar^2 \Gamma}\left(\tilde{\mathbf{E}}_{\mathrm{probe}} \cdot \overline{\mathbf{D}}\right) \times \mathbf{E}_{\mathrm{pu}}^{\mathrm{int}}(t) - \frac{e^3}{\varepsilon_0 \hbar^2 \Gamma}\left(\tilde{\mathbf{E}}_{\mathrm{pu}}(t) \cdot \overline{\mathbf{D}}\right) \times \tilde{\mathbf{L}} \cdot \left(\mathbf{E}_{\mathrm{probe}} - \mathbf{L} \cdot \frac{\mathbf{P}_{\mathrm{c,probe}}^{\mathrm{int}}}{\varepsilon_0}\right). \tag{13a}$$

$$\frac{d}{dt}\tilde{\mathbf{E}}_{\mathrm{probe}} + \Gamma \tilde{\mathbf{E}}_{\mathrm{probe}} = \Gamma \tilde{\mathbf{L}} \cdot \left(\mathbf{E}_{\mathrm{probe}} - \mathbf{L} \cdot \frac{\mathbf{P}_{\mathrm{c,probe}}^{\mathrm{int}}}{\varepsilon_0}\right). \tag{13b}$$

For convenience, I introduced the tensor $\tilde{\mathbf{L}} = \left(\mathbf{1} + (\varepsilon_\infty - 1)\mathbf{L}\right)^{-1}$.

The above equations determine a *linear* differential system for $\mathbf{P}_{\mathrm{c,probe}}^{\mathrm{int}}$ and $\tilde{\mathbf{E}}_{\mathrm{probe}}$ with time-dependent coefficients ($\mathbf{E}_{\mathrm{pu}}^{\mathrm{int}}(t), \tilde{\mathbf{E}}_{\mathrm{pu}}(t)$). Thus, the pump and the nonlinearity arising from the quantum geometry of the low-symmetry conductor effectively lead to time-crystalline behavior and optical-cycle modulation. The frequency of the driven time crystal is identical to the pump frequency ($\Omega$). For simplicity, I will identify the pump-generated fields $\mathbf{E}_{\mathrm{pu}}^{\mathrm{int}}(t), \tilde{\mathbf{E}}_{\mathrm{pu}}(t)$ with those calculated using the linear approximation [Eqs. (8) and (9)].

The component of the electric dipole induced by the probe interactions is:

$$\frac{\mathbf{p}_{\mathrm{dip,probe}}}{\varepsilon_0} = V\,\tilde{\mathbf{L}} \cdot \left[(\varepsilon_\infty - 1)\mathbf{E}_{\mathrm{probe}} + \frac{\mathbf{P}_{\mathrm{c,probe}}^{\mathrm{int}}}{\varepsilon_0}\right]. \tag{14}$$

Appendix C constructs a formal solution for the system (13) using a Floquet expansion of the fields. The analysis considers that the probe signal has a time-harmonic variation with frequency $\omega$.

## IV. Parametric Gain and Scattering Response

Next, I study the onset of parametric amplification and the pump-controlled scattering response of the plasmonic BCD nanoparticle.

Recent theoretical studies have shown that photonic time modulation can compensate plasmonic losses and enable parametric amplification in conducting and surface-plasmon systems [65, 66]. Furthermore, a substantial reduction of plasmonic losses due to parametric gain was recently demonstrated experimentally at terahertz frequencies using a plasmonic metamaterial [34]. More directly relevant here, time-crystalline media can strongly modify the power exchanged with localized sources [67] and produce near-field gain together with pronounced changes in far-field radiation [68, 69]. Here, rather than considering the effect of the time modulation on an external emitter, I show how optical pumping of a finite BCD nanoparticle can tailor its own scattering and absorption response. In particular, I demonstrate that its extinction cross section can become negative under sufficiently strong pumping.

### *A. Parametric gain*

In the remainder of the article, I assume for simplicity that the pump field is aligned with the $z$ axis. For a spheroidal BiTeI nanoparticle aligned along the same axis, this orientation maximizes the nonlinear response. In addition, it has an important advantage: due to the structure of the BCD tensor of BiTeI, the second-order nonlinear current generated by the pump alone vanishes identically. Consequently, in this configuration, the pump-induced interior fields given by Eqs.

(8) and (9) are exact within the quasistatic approximation. Furthermore, the complex amplitude of the pump field reduces to:

$$\mathbf{E}_{\mathrm{pu}0}^{\mathrm{int}} = \frac{1}{1+\left(\varepsilon\left(\Omega\right)-1\right)L_{zz}}\mathbf{E}_{\mathrm{pu}0}, \qquad \text{with } \mathbf{E}_{\mathrm{pu}0} = E_{\mathrm{pu}0}\hat{\mathbf{z}}\,. \tag{15}$$

This equation reveals something useful: the induced pump field is controlled by the depolarization factor $L_{zz}$, which depends on the geometry of the spheroidal nanoparticle. Since at the pump frequency the Drude permittivity is on the order of $\varepsilon_\infty$, it follows that shapes with a larger $L_{zz}$ effectively screen part of the incident field and weaken the nonlinear effects. Thus, elongated prolate spheroidal nanoparticles can provide stronger responses due to their small $L_{zz}$ ($L_{zz} \approx 0$), as compared with spherical particles ($L_{zz} = 1/3$).

Furthermore, for a pump field along *z*, Eq. (13a) reduces to:

$$\frac{d}{dt}\frac{\mathbf{P}_{\mathrm{c,probe}}^{\mathrm{int}}}{\varepsilon_0} = \frac{\varepsilon_\infty \omega_{\mathrm{ENZ}}^2}{\Gamma}\tilde{\mathbf{E}}_{\mathrm{probe}} - \frac{e^3 D_{\mathrm{B}}}{\varepsilon_0 \hbar^2 \Gamma}\left(\hat{\mathbf{z}}\times\tilde{\mathbf{E}}_{\mathrm{probe}}\right)\times\mathbf{E}_{\mathrm{pu}}^{\mathrm{int}}\left(t\right). \tag{16}$$

Equation (16), together with Eq. (13b), shows that a probe polarized parallel to the pump, i.e., aligned along *z*, is insensitive to the nonlinearity.

To determine the conditions for parametric instability, I set the incident probe field equal to zero ($\mathbf{E}_{\mathrm{probe}} = 0$) in Eq. (13b). Then, the transverse components of $\tilde{\mathbf{E}}_{\mathrm{probe}}, \mathbf{P}_{\mathrm{c,probe}}^{\mathrm{int}}$ (say, along *x*), satisfy the homogeneous scalar equations:

$$\frac{d}{dt}P_{\mathrm{c}}^{\mathrm{int}} = \varepsilon_0\left(\frac{\varepsilon_\infty \omega_{\mathrm{ENZ}}^2}{\Gamma} - \frac{e^3 D_{\mathrm{B}}}{\varepsilon_0 \hbar^2 \Gamma}E_{\mathrm{pu}}^{\mathrm{int}}\left(t\right)\right)\tilde{E}, \tag{17a}$$

$$\frac{d}{dt}\tilde{E} + \Gamma\tilde{E} = -\Gamma\frac{L_t}{1+L_t\left(\varepsilon_\infty - 1\right)}\frac{P_{\mathrm{c}}^{\mathrm{int}}}{\varepsilon_0}, \tag{17b}$$

with $\mathbf{E}_{\text{pu}}^{\text{int}}(t) = E_{\text{pu}}^{\text{int}}(t)\hat{\mathbf{z}}$. For notational simplicity, the "probe" subscripts have been omitted. Note also that the chosen pump does not generate transverse interior fields.

Eliminating $P_{\text{c}}^{\text{int}}$ by combining Eqs. (17a)-(17b) yields the following homogeneous equation for $\tilde{E}$

$$\frac{d^2}{dt^2}\tilde{E} + \Gamma\frac{d\tilde{E}}{dt} + \omega_0^2\left(1+\delta_{\text{m}}(t)\right)\tilde{E} = 0. \tag{18}$$

Here,

$$\omega_0 = \sqrt{\frac{L_t\varepsilon_\infty}{1+L_t(\varepsilon_\infty - 1)}}\omega_{\text{ENZ}}, \tag{19}$$

is the transverse resonance frequency of the undamped spheroidal nanoparticle, whereas

$$\delta_{\text{m}}(t) = \frac{-e^3 D_{\text{B}}}{\varepsilon_0\varepsilon_\infty\hbar^2\omega_{\text{ENZ}}^2}E_{\text{pu}}^{\text{int}}(t), \tag{20}$$

describes the effect of the pump modulation. For the material parameters considered here, the resonance frequency remains close to the ENZ point.

Clearly, Eq. (18) has the form of a damped Mathieu equation. For relatively weak modulations and $\Omega \sim 2\omega_0'$, the corresponding Floquet quasi-frequency is given by [65, 70]:

$$\omega_{\text{res}} = \omega_0' + i\omega_0'', \qquad \text{with} \qquad \omega_0'' \approx -\frac{\Gamma}{2} + \omega_0' \ \text{Re}\left\{\sqrt{\left(\frac{\delta_{\text{M}}}{4}\frac{\omega_0^2}{\omega_0'^2}\right)^2 - \left(\frac{\Omega}{2\omega_0'} - 1\right)^2}\right\}. \tag{21}$$

with $\omega_0' = \sqrt{\omega_0^2 - \left(\frac{\Gamma}{2}\right)^2}$. Here, $\delta_{\text{M}}$ is the peak value of the modulation function $\delta_{\text{m}}(t)$. Using Eq. (8), it can be explicitly written as:

$$\delta_{\text{M}} = \frac{e^3|D_{\text{B}}|}{\varepsilon_0\varepsilon_\infty\hbar^2\omega_{\text{ENZ}}^2}\frac{|E_{\text{pu}0}|}{|1+(\varepsilon(\Omega)-1)L_{zz}|}. \tag{22}$$

A parametric instability requires that $\omega_0'' > 0$. For the optimal modulation frequency, $\Omega = 2\omega_0'$, this requires that the peak modulation depth obeys $\delta_{\mathrm{M}} \geq \frac{2\Gamma\omega_0'}{\omega_0^2} \approx \frac{2\Gamma}{\omega_{\mathrm{ENZ}}}$. The approximate identity assumes that $\frac{2\Gamma}{\omega_{\mathrm{ENZ}}}$ is small. The parametric gain condition yields the following threshold for the peak amplitude of the pump field:

$$\begin{aligned} E_{\mathrm{pu,th}} &= \frac{\varepsilon_0\varepsilon_\infty\hbar^2\omega_{\mathrm{ENZ}}^2}{e^3\left|D_{\mathrm{B}}\right|}\left|1+\left(\varepsilon\left(\Omega\right)-1\right)L_{zz}\right|\frac{2\Gamma\omega_0'}{\omega_0^2} \\ &\approx \frac{2\varepsilon_0\varepsilon_\infty\hbar^2\omega_{\mathrm{ENZ}}\Gamma}{e^3\left|D_{\mathrm{B}}\right|}\left|1+\left(\varepsilon\left(\Omega\right)-1\right)L_{zz}\right|. \end{aligned} \tag{23}$$

Figure 2b depicts the threshold pump field as a function of the depolarization factor $L_{zz}$ for $\omega_{\mathrm{ENZ}} = 2\pi \times 12.5\mathrm{THz}$. The term $L_{zz}$ is a function of the ratio of the two main radii of the spheroid ($R_t / R_z$), represented in Fig. 2a [64]. One can see that elongated spheroidal particles, with $L_{zz}$ approaching zero, require a weaker incident pump at the onset of the parametric instability. As already discussed, this occurs because geometries with larger $L_{zz}$ more effectively screen the incident pump field, reducing its penetration into the nanoparticle.

The pump field threshold for an elongated nanoparticle ($R_z / R_t = 3.2$) with $L_{zz} = 0.1$ is on the order of $2\times10^8$ V/m=200 MV/m. Such peak fields are within the range commonly accessed with short optical pulses. For comparison, ultrafast experiments in TCOs have employed peak fields about 20 times larger than those considered here [26]. Moreover, the pump pulse need not be extremely short. A pulse containing one hundred optical cycles has a duration of only a few picoseconds, while still providing sufficiently many modulation periods for the Floquet instability to develop. The required pulse energy can remain relatively small for a tightly focused beam. For example, a diffraction-limited pulse containing about one hundred optical cycles corresponds to an energy on the order of tens of nJ at the instability threshold. This is several orders of magnitude smaller than the pulse energies used in Ref. [26] at optical frequencies.

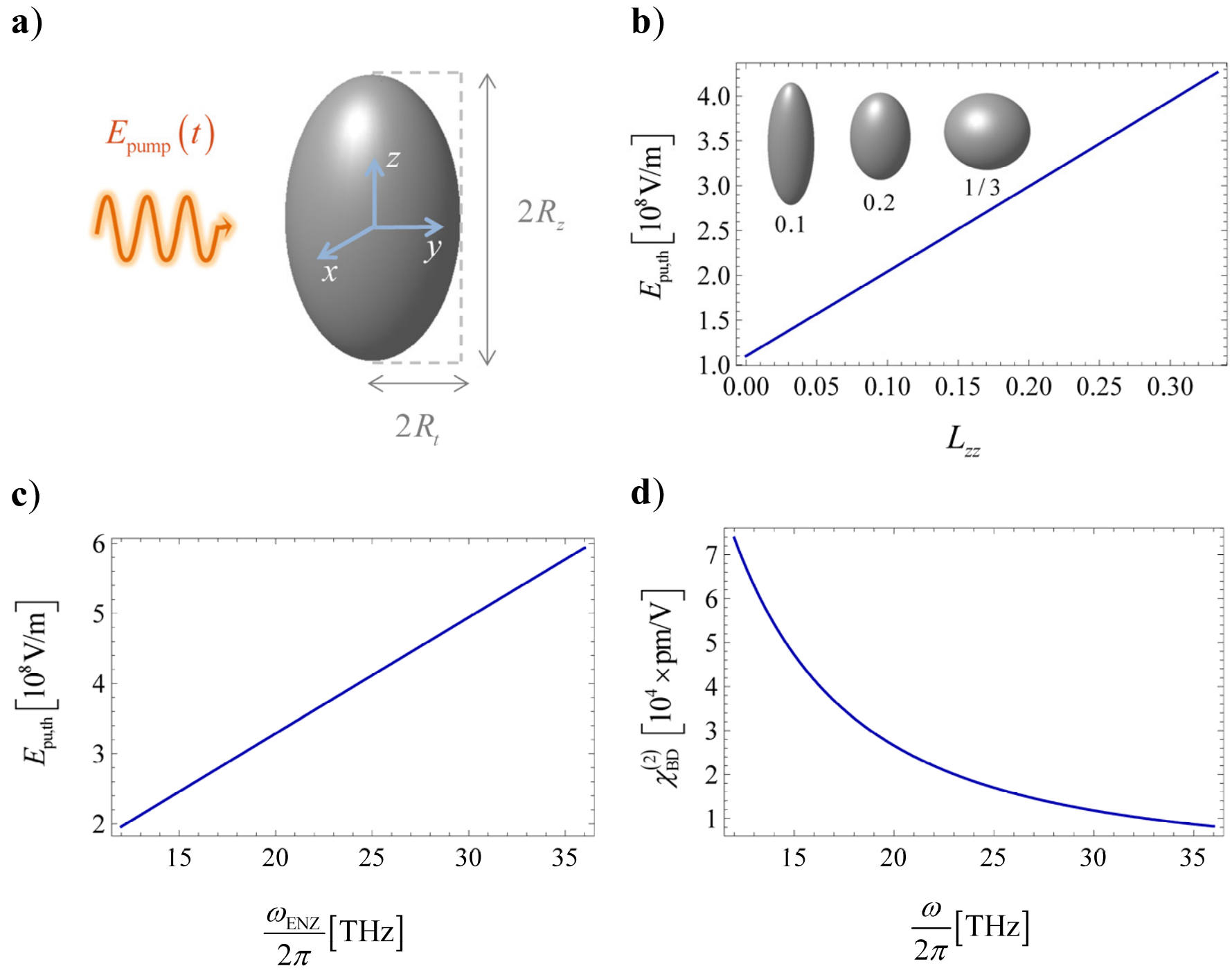

**Fig. 2** **(a)** Sketch of the geometry of the spheroidal nanoparticle. The spheroid symmetry axis is aligned with the $z$-axis, i.e., with the direction of the pump field. **(b)** Pump field at the threshold of the parametric instability as a function of the depolarization factor $L_{zz}$ for $\omega_{\text{ENZ}} = 2\pi \times 12.5\,\text{THz}$. The insets illustrate the shape of the spheroidal particle for different values of the depolarization factor. **(c)** Pump-field threshold as a function of the ENZ frequency for $L_{zz} = 0.1$. **(d)** Effective second-order susceptibility of BiTeI as a function of frequency. In all panels, BiTeI is modeled with $\varepsilon_\infty = 13$, $\Gamma = 2\pi \times 3.5\,\text{THz}$, and $D_{\text{B}} = 0.01$.

Figure 2c illustrates the impact of changing the ENZ crossing point on the threshold field. Note that the ENZ frequency depends on the doping concentration, and can thus be tuned through doping. The range of $\omega_{\text{ENZ}}$ in the plot is estimated to correspond to a chemical-potential window on the order of 0.02-0.18 eV where the Berry curvature dipole varies relatively slowly with the electron concentration [61]. As seen, the nonlinear response is stronger at the lower end of the $\omega_{\rm ENZ}$ scale. At first, this effect is counterintuitive: as the collision rate $\Gamma$ is assumed constant in the plot of Fig. 2c, the quality factor of the plasmonic resonance, $Q = \omega_{\rm ENZ} / \Gamma$, decreases toward the lower end of the scale, in apparent contradiction with the fact that it corresponds to the minimum pump field for a parametric instability.

The explanation for this behavior is that the nonlinear effects are greatly enhanced at lower frequencies. In fact, one can estimate from the BCD-Drude transport equations (2)-(3) that the effective second-order susceptibility associated with the Berry-curvature dipole is [33] (see also Appendix A):

$$\chi_{\rm BD}^{(2)} = D_{\rm B} \frac{1}{\omega(\omega + i\Gamma)} \frac{e^3}{\varepsilon_0 \hbar^2} . \tag{24}$$

The corresponding induced tensorial permittivity change is $\Delta\varepsilon_{\rm BD} \sim \varepsilon_0 \chi_{\rm BD}^{(2)} E_{\rm pump}$. The Drude-like factor $1/\left[\omega(\omega + i\Gamma)\right]$ describes the fact that at lower frequencies the carriers can be displaced over larger distances during one optical cycle, resulting in a stronger $\chi_{\rm BD}^{(2)}$.

Hence, provided the quality factor $Q = \omega_{\rm ENZ} / \Gamma$ is at least a few times larger than unity, lowering the ENZ frequency by a factor of 2 boosts the effective $\chi_{\rm BD}^{(2)}$ by a factor of 4, which is enough to compensate for the corresponding reduction of the quality factor. As a result, the instability threshold decreases approximately linearly with the ENZ frequency.

Figure 2d depicts the amplitude of $\chi_{\rm BD}^{(2)}$ as a function of frequency. As seen, $\chi_{\rm BD}^{(2)}$ can reach enormous values on the order of $7 \times 10^4$ pm/V. In the conventional susceptibility-tensor notation, the BCD response shown in Fig. 2d corresponds to $\chi_{xxz}^{(2)} = \chi_{xzx}^{(2)} = \chi_{\rm BD}^{(2)} / 2$. For comparison, GaAs has a second-order susceptibility $\left|\chi_{xyz}^{(2)}\right| \approx 400\text{pm/V}$ for photon energies slightly below the direct interband edge [60]. Thus, the BCD nonlinearity of BiTeI can exceed the near-resonant second-order response of GaAs by about two orders of magnitude. Strong second-order optical nonlinearities have also been observed experimentally in BiTeI at optical frequencies [59], although these originate from a different microscopic mechanism.

It is instructive to write the threshold pump field in terms of the effective $\chi_{\text{BD}}^{(2)}$ susceptibility evaluated at the plasmonic resonance ($\omega \approx \omega_{\text{ENZ}}$). Using $\left|\chi_{\text{BD}}^{(2)}\right| \approx |D_{\text{B}}| \frac{1}{\omega_{\text{ENZ}}^2} \frac{e^3}{\varepsilon_0 \hbar^2}$ and the approximations $\omega_0 \approx \omega_0' \approx \omega_{\text{ENZ}}$, one finds from Eq. (23) that:

$$E_{\text{pu,th}} \approx \frac{\varepsilon_\infty}{\left|\chi_{\text{BD}}^{(2)}\right|} \frac{1}{Q} \times 2\left|1 + \left(\varepsilon(\Omega) - 1\right) L_{zz}\right|, \tag{25}$$

with $Q = \omega_{\text{ENZ}} / \Gamma$ the plasmonic resonance quality factor. Even though this result is specific to the plasmonic resonator, its main scaling is shared by a broader class of systems [71]. Specifically, the threshold field is inversely proportional to the quality factor of the relevant resonant mode and to the strength of the nonlinearity, measured by the second-order susceptibility. In addition, it is also inversely proportional to the term $1/\left|1 + \left(\varepsilon(\Omega) - 1\right) L_{zz}\right|$, which determines the local-field coupling of the incident pump to the nanoparticle.

Equation (25) reinforces the strengths and the weaknesses of systems relying on conducting materials: their strong nonlinear response contributes to lowering the threshold pump field, whereas their poor quality factor contributes to increasing it. Compared to dielectric-resonator architectures [71, 72], such as GaAs-based resonators, the large BCD susceptibility of BiTeI can compensate for quality factors almost two orders of magnitude smaller than those attainable with nonlinear dielectric resonators. Furthermore, dielectric-based resonators typically rely on Mie-type modes, and the nonlinear interaction may therefore be penalized by imperfect spatial-overlap factors and symmetry constraints [71, 72]. In contrast, for the subwavelength plasmonic resonator considered here, the relevant interior field structures are spatially uniform, resulting in an essentially ideal spatial overlap.

To conclude this subsection, I note that the pumped BCD nanoparticle may be regarded as an optically pumped spaser [73].

### *B. Effective scattering response and negative extinction*

As seen in Sect. IV.A, observing parametric amplification may be challenging because the pump pulse may need to be limited to a few hundred optical cycles to maintain a reasonable pulse energy. A perhaps more practical experimental route is to probe the scattering response of the nanoparticle using a weak signal at frequency $\omega$. The most direct signature of the effect of the pump on the probe is frequency mixing, as the interaction of the two waves mediated by the nanoparticle naturally generates frequency components at $\omega \pm \Omega$, i.e., Floquet sidebands. However, here I am mostly interested in the effect of the pump on the main harmonic with frequency $\omega$.

In principle, such an effect is comparatively weaker than the frequency-mixing effect, as it results from the coupling of the main harmonic at $\omega$ to the Floquet sidebands at $\omega \pm \Omega$, and their subsequent coupling back to the main harmonic. Thus, it is a second-order effect in the pump-induced modulation. This should be contrasted with the behavior of a $\chi^{(3)}$ nonlinearity, which contributes to the main harmonic already at first order in the nonlinear coupling. From a different perspective, the effect of the $\chi^{(2)}$ modulation on the material response tends to average out over an optical cycle, because the modulation continuously oscillates about the linear response. Thus, a $\chi^{(2)}$ modulation may more weakly affect observables associated with the main harmonic, such as scattering or absorption. This property is characteristic of the time-crystalline response. It should be noted that the effective $\chi^{(3)}$ nonlinearity typically exploited in TCOs [25-28], which is associated with an abrupt change in the material response followed by a relatively slow relaxation, is in this respect well suited to producing large changes in the optical response. Indeed, the material is effectively switched from one state to another, and the modified response persists for many optical cycles.

To characterize the response at the main harmonic, I introduce an effective polarizability tensor, $\overline{\boldsymbol{\alpha}}_{\rm ef}(\omega)$, defined such that it links the main harmonic of the probe-induced electric dipole moment, $\mathbf{p}_{\rm dip,probe}$, with the probe field $\mathbf{E}_{\rm probe}(t)=\mathbf{E}_{0\omega}e^{-i\omega t}$. Specifically,

$$\left(\mathbf{p}_{\rm dip,probe}\right)_{\omega}=\varepsilon_0\overline{\boldsymbol{\alpha}}_{\rm ef}(\omega)\cdot\mathbf{E}_{0\omega}. \tag{26}$$

Thus, $\overline{\boldsymbol{\alpha}}_{\rm ef}$ provides a homogenized description of the effect of the pump modulation at the probe frequency. The tensor $\overline{\boldsymbol{\alpha}}_{\rm ef}$ can be found by numerically solving the linearized system of equations (13) for three different orientations of $\mathbf{E}_{0\omega}$, typically aligned with the coordinate axes. Note that the main harmonic of $\mathbf{p}_{\rm dip,probe}$ can be related to the main harmonic of $\mathbf{E}_{\rm probe}$ and $\mathbf{P}_{\rm c,probe}^{\rm int}$ using Eq. (14). Thus, using this relation and the numerical solution described in Appendix C, it is straightforward to obtain $\overline{\boldsymbol{\alpha}}_{\rm ef}$.

Here, I consider the same geometrical configuration as in Sect. IV: the symmetry axis of the spheroidal particle and the pump field are both aligned with the *z* axis [Fig. 2a]. Furthermore, in all the examples I assume that the depolarization factor is $L_{zz}=0.1$, the ENZ frequency is $\omega_{\rm ENZ}=2\pi\times12.5\,{\rm THz}$, and that the pump frequency is $\Omega=2\omega_0'\approx2\pi\times23.6\,{\rm THz}$. The conclusions remain qualitatively similar for other parameter choices. For the parameters considered here, the parametric-instability threshold is $E_{\text{pu,th}} = 2\times 10^8\ \text{V/m}$.

Due to the rotational symmetry of the adopted model about the $z$ axis and the alignment of the pump field along this axis, the effective polarizability has a uniaxial structure of the form $\bar{\boldsymbol{\alpha}}_{\text{ef}}(\omega) = \alpha_{\text{ef},t}(\omega)(\hat{\mathbf{x}}\otimes\hat{\mathbf{x}} + \hat{\mathbf{y}}\otimes\hat{\mathbf{y}}) + \alpha_{\text{ef},zz}(\omega)\hat{\mathbf{z}}\otimes\hat{\mathbf{z}}$. The $zz$ component of the tensor is independent of the pump parameters. In fact, as already discussed, a probe aligned along $z$ is insensitive to the pump-induced nonlinearity. Hence, in the rest of this subsection I will focus exclusively on the transverse component $\alpha_{\text{ef},t}(\omega)$.

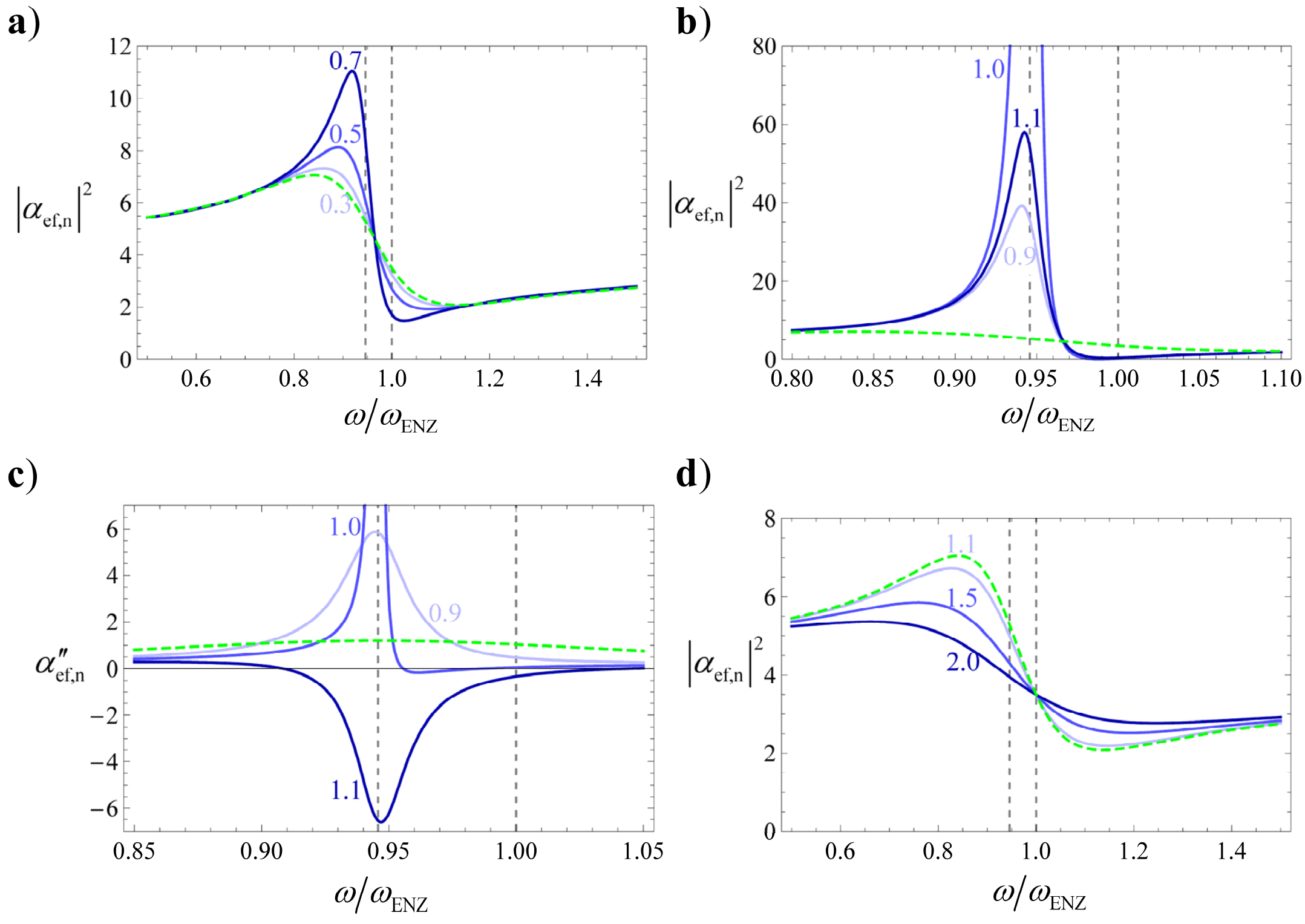

**Fig. 3** **(a)-(b)** Normalized scattering strength as a function of the probe frequency for different pump-field amplitudes. The simulations use $\Gamma = 2\pi\times 3.5\,\text{THz}$, $\omega_{\text{ENZ}} = 2\pi\times 12.5\,\text{THz}$, and $L_{zz} = 0.1$. The labels indicate the peak pump-field amplitude normalized to the instability threshold, $|E_{\text{pu}}|/E_{\text{pu,th}}$, with colors matching the corresponding curves. The green dashed curve represents the linear response. **(c)** Normalized extinction strength as a function of the probe frequency for different pump-field amplitudes; all other parameters and conventions are as in panels (a)-(b). **(d)** Same as panels (a)-(b), but for a weak pump field, ($\left|E_{\rm pu}\right|/E_{\rm pu,th}=0.1$) and different values of $\Gamma$. The labels indicate $\Gamma/\left(2\pi\times3.5\,{\rm THz}\right)$. The green dashed curve shows the linear response for the baseline $\Gamma=2\pi\times3.5\,{\rm THz}$. In all panels, the pump frequency is $\Omega=2\omega_0'=2\pi\times23.6\,{\rm THz}$, $\varepsilon_\infty=13$ and $D_{\rm B}=0.01$; the two vertical dashed lines mark $\omega=\omega_0'$ and $\omega=\omega_{\rm ENZ}$.

The response of the spheroidal nanoparticle to the probe field can be conveniently characterized through the extinction power ($P_{\rm ext}$) and the scattered power ($P_{\rm scat}$) at the main harmonic:

$$P_{\rm ext}=\frac{\omega}{c}{\rm Im}\left\{\alpha_{{\rm ef},t}\right\}S_0\,,\qquad\qquad P_{\rm scat}=\frac{1}{6\pi}\left(\frac{\omega}{c}\right)^4\left|\alpha_{{\rm ef},t}\right|^2S_0\,,\tag{27}$$

where $S_0=\dfrac{\left|\mathbf{E}_{0\omega}\right|^2}{2\eta_0}$ is the intensity of the incoming probe wave, and $\eta_0$ is the vacuum impedance. It is implicit that the probe field is polarized in the *xy* plane. I recall that the extinction can be expressed as the sum of the scattered and absorbed powers: $P_{\rm ext}=P_{\rm scat}+P_{\rm abs}$. Physically, it characterizes the power that is "removed" from the incoming probe beam (along a narrow angular range that contains the forward scattering direction), after it interacts with the scatterer [75].

The effect of the pump on both $P_{\rm scat}$ and $P_{\rm ext}$ is governed by the effective polarizability. Thus, I characterize the scattering and extinction strengths by the dimensionless parameters, $\left|\alpha_{\rm ef,n}\right|^2$ and ${\rm Im}\left\{\alpha_{\rm ef,n}\right\}\equiv\alpha''_{\rm ef,n}$, respectively, with $\alpha_{\rm ef,n}=\alpha_{{\rm ef},t}/V$, where $V$ is the volume of the spheroidal nanoparticle.

Figures 3a and 3b illustrate the scattering strength for different pump amplitudes, determined by the values of $\left|E_{\text{pu}}\right|/E_{\text{pu,th}}$ shown in the labels next to the blue curves. Note that since the frequency-dependent term $\left(\frac{\omega}{c}\right)^4$ was omitted from the normalized parameter $\left|\alpha_{\text{ef,n}}\right|^2$, the actual frequency dependence of $P_{\text{scat}}$ is different from what is shown in Fig. 3. The key objective here is to understand how the scattering changes in the presence of the pump, relative to the baseline linear response (the green dashed line in all plots). As seen in Fig. 3a, for $\left|E_{\text{pu}}\right|/E_{\text{pu,th}} = 0.3$ the pump is already strong enough to produce a visible variation of the scattering strength near the transverse plasmonic resonance, $\omega \sim \omega_0'$. For stronger pumps, approaching the instability threshold, the effect becomes increasingly pronounced and is particularly visible in and around the interval $\omega_0' < \omega < \omega_{\text{ENZ}}$ . In particular, the scattering strength is greatly enhanced at the plasmonic resonance $\omega = \omega_0'$ and strongly depleted near the ENZ point $\omega = \omega_{\text{ENZ}}$.

Furthermore, exactly at the threshold the effective polarizability has a pole, and the scattering strength $\left|\alpha_{\text{ef,n}}\right|^2$ diverges to infinity, as illustrated in Fig. 3b. This behavior is consistent with the fact that a pole crosses the real-frequency axis at the instability threshold, also causing the steady-state response to become ill-defined. For pump fields stronger than the threshold (see the curve $\left|E_{\text{pu}}\right|/E_{\text{pu,th}} = 1.1$ in Fig. 3b), the scattering strength becomes finite again. For a finite-duration pump, this response may remain physically relevant if the instability growth time is longer than the pump duration [74].

Next, I turn to the extinction strength, represented in Figure 3c. Similar to the scattering strength, it is strongly enhanced at the instability threshold. This is expected due to the link between the extinction and the scattering powers. Remarkably, Fig. 3c reveals that for pump fields sufficiently close to the instability threshold the extinction strength can become negative, in other words $P_{\text{ext}} < 0$. Interestingly, as suggested by the curve with $|E_{\text{pu}}|/E_{\text{pu,th}} = 1.0$ in Fig. 3c, slightly above the plasmonic resonance, negative extinction may also occur in the stable regime ($|E_{\text{pu}}|/E_{\text{pu,th}} < 1.0$).

As the scattering power is always positive, a negative extinction implies that the absorbed power ($P_{\text{abs}}$) must also be negative. In other words, the homogenized particle behaves as a gain system, and rather than absorbing energy from the probe, it emits energy into the far zone. Since the extinction determines the energy extracted specifically from the incident probe beam [75], it follows that when the extinction is negative the nanoparticle effectively amplifies the probe beam, so that its intensity increases after interacting with the particle. Additionally, in this regime, the net scattered radiation (i.e., the radiation propagating into angular channels that exclude the main forward beam direction) also originates from the gain of the nanoparticle. Related ideas have been discussed in previous work in different contexts [76, 77].

In all the analyses so far, I have assumed that the collision rate in the material is not affected by the optical pump. However, it is likely, considering the behavior of TCOs [16-18] and the known temperature dependence of the collision rate of BiTeI [62], that $\Gamma$ is sensitive to the pump intensity. It is thus natural to wonder whether changes in $\Gamma$ relative to the linear baseline can also have an impact on the effective polarizability. To investigate this, I plot in Fig. 3d the normalized scattering strength for different values of $\Gamma$. The normalized pump amplitude for all the blue curves is only $|E_{\text{pu}}|/E_{\text{pu,th}} = 0.1$. As expected, increasing $\Gamma$ flattens the resonant behavior around $\omega \sim \omega_0'$, in contrast to the direct effect of parametric pumping, which sharpens the resonance (Figs. 3a and 3b). Hence, in a practical scenario there will likely be a competition between the two effects.

## V. Summary and Discussion

In summary, I have presented a study of the nonlinear response of low-symmetry conductors and demonstrated that these systems are natural platforms for observing a time-crystalline response in pump-probe experiments. In particular, I have demonstrated that, within the semiclassical approximation, the nonlinearity is exactly described by a time-domain BCD-Drude formalism that compactly aggregates all the relevant nonlinear couplings. In addition, I have discussed the intrinsic tradeoff between the strength of the nonlinearity and operation below the electronic interband threshold. This tradeoff was formalized through a bound on the Berry-curvature dipole that scales inversely with the square of the relevant minimum interband energy separation. Then, I developed a master system of equations that characterizes the electrodynamics of a subwavelength BCD nanoparticle in a pump-probe configuration. Based on this model, I studied the conditions for parametric amplification in a BiTeI nanoparticle and found that the required pump amplitude is, in principle, within experimental reach. Furthermore, I studied the influence of the pump on the response at the main harmonic of the probe field using an effective polarizability. The analysis shows that field strengths comparable to the instability threshold can produce strong changes in the scattering and extinction of the nanoparticle, which can in principle be experimentally characterized. Finally, I uncovered a regime of negative extinction, in which the incident probe beam gains energy from the interaction with the nanoparticle.

In the system studied here, the large BCD nonlinearity comes together with the relatively poor quality factor imposed by Drude losses. This tradeoff is not necessarily fundamental. For example, one may envision other resonant structures in which the confinement does not arise directly from the plasmonic response of the BCD particle. In that case, the low-symmetry conductor would instead primarily provide the nonlinear coupling to a high-Q resonance supported by a low-loss structure. This strategy could retain a substantial fraction of the strong BCD response while avoiding, at least in part, the loss penalty associated with a plasmonic resonance. Such configurations may remain relevant even in spectral regions where the nonlinear response includes corrections beyond the simple BCD-Drude description. Exploring such configurations, and determining the extent to which the nonlinear and dissipative parts of the response can be independently engineered, is an interesting direction for future work.

**Acknowledgements:** This work is supported in part by the Institution of Engineering and Technology (IET), by the Simons Foundation (award SFI-MPS-EWP-00008530-10), and by national funds through FCT – Fundação para a Ciência e a Tecnologia, I.P., and, when eligible, co-funded by EU funds under project/support UID/50008/2025 – Instituto de Telecomunicações, with DOI identifier https://doi.org/10.54499/UID/50008/2025.

## Appendix A: Equivalence between the time-domain nonlinear model and the standard semiclassical formalism

Here, I show that the time-domain BCD-Drude model adopted in the main text is exactly equivalent to the standard semiclassical description of the BCD response.

The nonlinear optical response of a crystalline material may be calculated from a microscopic Kubo-type response theory. A particularly convenient formulation is the length-gauge density-matrix method of Aversa and Sipe [52]. The formalism treats interband transitions and intraband motion in a gauge-consistent manner. Aversa and Sipe applied the method to clean and cold semiconductors within the independent-particle approximation. Subsequent works extended and adapted the length-gauge framework to doped and conducting systems [53-55].

A useful decomposition of the second-order conductivity, introduced by Matsyshyn and Sodemann, separates the response of a time-reversal-invariant conductor into Berry-curvature-dipole, injection, and residual interband contributions [56]. For frequencies well below the relevant interband transition frequencies, it has been demonstrated that the leading metallic response is the Berry-curvature-dipole contribution [55-56]. Specifically, for two field components with frequencies $\omega_1$ and $\omega_2$, the resulting current at $\omega_1+\omega_2$ can be expressed as [55-56]:

$$\mathbf{j}^{(2)}_{\omega_1+\omega_2} = -\frac{e^3}{\hbar^2}\frac{1}{\Gamma - i\omega_2}\left[\mathbf{E}_{\omega_2}(t)\cdot\overline{\mathbf{D}}\right]\times\mathbf{E}_{\omega_1}(t) - \frac{e^3}{\hbar^2}\frac{1}{\Gamma - i\omega_1}\left[\mathbf{E}_{\omega_1}(t)\cdot\overline{\mathbf{D}}\right]\times\mathbf{E}_{\omega_2}(t). \tag{A1}$$

where $\overline{\mathbf{D}}$ is the Berry-curvature-dipole tensor. It is implicit that the time dependence of $\mathbf{E}_{\omega_i}(t)$ is of the form $e^{-i\omega_i t}$. The two terms describe the two possible orderings of the field interactions. The above result coincides with that obtained using a Boltzmann description, and includes as particular cases the examples studied in Refs. [35, 43, 44].

Remarkably, the complete frequency dependence of this semiclassical BCD response can be aggregated into the time-domain model discussed in the main text. Indeed, for a generic electric field, $\mathbf{E}(t) = \sum_i \mathbf{E}_{\omega_i}(t)$, the total second-order current can be expressed as:

$$\mathbf{j}^{(2)}(t) = \sum_{\omega_i,\omega_j} -\frac{e^3}{\hbar^2}\frac{1}{\Gamma - i\omega_j}\left[\mathbf{E}_{\omega_j}(t)\cdot\overline{\mathbf{D}}\right]\times\mathbf{E}_{\omega_i}(t). \tag{A2}$$

This coincides with the current density obtained by combining Eqs. (2)-(3) of the main text. Thus, this representation unifies rectification, harmonic generation, frequency mixing, and transient effects in a single dynamical constitutive law. The approach is particularly convenient for coupling the electronic response directly to Maxwell's equations.

## Appendix B: Bound for the Berry Curvature Dipole

In this Appendix, I derive a simple bound on the magnitude of the components of the Berry curvature dipole tensor $\overline{\mathbf{D}}$ of a three-dimensional material. The element $D_{ij}$ of the tensor can be written explicitly in terms of the Berry curvature as:

$$D_{ij} = \frac{1}{(2\pi)^3} \iiint f_{\mathbf{k}} \partial_i \Omega_{\mathbf{k}j} d^3\mathbf{k}\,, \tag{B1}$$

Here, $f_{\mathbf{k}}$ denotes the Fermi-Dirac distribution, $\partial_i = \partial / \partial k_i$ is a derivative with respect to the *i*-th component of the wave vector and $\Omega_{\mathbf{k}j}$ is the *j*-th component of the Berry curvature. The integration is over the Brillouin zone.

The low-temperature limit is assumed so that $f_{\mathbf{k}}$ has a step-like variation, $f_{\mathbf{k}} = \Theta(\mu - E_{n\mathbf{k}})$ , where $\mu$ is the chemical potential, $E_{n\mathbf{k}}$ is the energy of the relevant (conduction or valence) band responsible for the electronic transport, and $\Theta$ is the Heaviside step function.

Integrating by parts Eq. (B1), the integration region can be reduced to the Fermi surface, $\mu = E_{n\mathbf{k}}$ :

$$D_{ij} = \frac{1}{(2\pi)^3} \iint_{E_{n\mathbf{k}}=\mu} \Omega_{\mathbf{k}j} \frac{\partial_i E_{n\mathbf{k}}}{\left|\nabla_{\mathbf{k}} E_{n\mathbf{k}}\right|} ds\,. \tag{B2}$$

I used the relation $\partial_i f_{\mathbf{k}} = -\partial_i E_{n\mathbf{k}} \delta(E_{n\mathbf{k}} - \mu)$. The above expression leads to a natural bound on the magnitude of the Berry curvature dipole components in terms of the maximum of the Berry curvature itself:

$$\left|D_{ij}\right| \le \frac{S_F}{(2\pi)^3} \max_{E_{n\mathbf{k}}=\mu} \left|\Omega_{\mathbf{k}j}\right|, \tag{B3}$$

where $S_F$ is the area of the Fermi surface.

On the other hand, the Berry curvature of a given band can be expressed in terms of the Bloch eigenstates $|u_{m\mathbf{k}}\rangle$ as [78]:

$$\Omega_{\mathbf{k}j} = i\frac{\varepsilon_{jrs}\hbar^2}{2}\sum_{m\mathbf{k}\neq n\mathbf{k}}\left[\langle u_{n\mathbf{k}}|\hat{v}_{\mathbf{k},r}|u_{m\mathbf{k}}\rangle\langle u_{m\mathbf{k}}|\hat{v}_{\mathbf{k},s}|u_{n\mathbf{k}}\rangle - \langle u_{m\mathbf{k}}|\hat{v}_{\mathbf{k},r}|u_{n\mathbf{k}}\rangle\langle u_{n\mathbf{k}}|\hat{v}_{\mathbf{k},s}|u_{m\mathbf{k}}\rangle\right]\frac{1}{\left[E_{mn,\mathbf{k}}\right]^2}, \quad \text{(B4)}$$

where $\hat{\mathbf{v}}_\mathbf{k} = \partial_\mathbf{k}\hat{H}_\mathbf{k}/\hbar$ is the velocity operator, $\hat{H}_\mathbf{k}$ is the Bloch Hamiltonian, $E_{mn,\mathbf{k}} = E_{m\mathbf{k}} - E_{n\mathbf{k}}$, and $\varepsilon_{jrs}$ is the Levi-Civita symbol. Note that $\hat{H}_\mathbf{k}|u_{m\mathbf{k}}\rangle = E_{m\mathbf{k}}|u_{m\mathbf{k}}\rangle$ and the eigenfunctions are normalized such that $\|u_{m\mathbf{k}}\| = 1$.

Similar to Ref. [79], I express the Berry curvature as:

$$\Omega_{\mathbf{k}j} = i\frac{\varepsilon_{jrs}\hbar^2}{2}\left[\langle u_{n\mathbf{k}}|\mathbf{A}_{n\mathbf{k}}^{rs}|u_{n\mathbf{k}}\rangle - \langle u_{n\mathbf{k}}|\mathbf{A}_{n\mathbf{k}}^{sr}|u_{n\mathbf{k}}\rangle\right], \qquad \text{with} \qquad \text{(B5a)}$$

$$\mathbf{A}_{n\mathbf{k}}^{rs} = \hat{v}_{\mathbf{k},r}G_{\mathbf{k},n}\hat{v}_{\mathbf{k},s} \quad \text{and} \quad G_{\mathbf{k},n} = \sum_{m\mathbf{k}\neq n\mathbf{k}}\frac{1}{\left[E_{mn,\mathbf{k}}\right]^2}|u_{m\mathbf{k}}\rangle\langle u_{m\mathbf{k}}|. \qquad \text{(B5b)}$$

Taking into account that $\left|\langle u_{n\mathbf{k}}|\mathbf{A}_{n\mathbf{k}}^{rs}|u_{n\mathbf{k}}\rangle\right| \leq \left\|\mathbf{A}_{n\mathbf{k}}^{rs}\right\|_\infty \|u_{n\mathbf{k}}\|^2$ and using the bound for the spectral norm $\|\mathbf{A}_1\mathbf{A}_2....\mathbf{A}_N\|_\infty \leq \|\mathbf{A}_1\|_\infty ...\|\mathbf{A}_N\|_\infty$, it follows that:

$$\left|\Omega_{\mathbf{k}j}\right| \leq \hbar^2\left|\varepsilon_{jrs}\right| \times \left\|\hat{v}_{\mathbf{k},r}\right\|_\infty \left\|\hat{v}_{\mathbf{k},s}\right\|_\infty \left\|G_{\mathbf{k},n}\right\|_\infty, \qquad \text{(B6)}$$

where I used $\|u_{n\mathbf{k}}\| = 1$. The spectral norm of $G_{\mathbf{k},n}$ can be bounded in terms of the minimum direct energy separation $\Delta$ [79]:

$$\left\|G_{\mathbf{k},n}\right\|_\infty \leq \frac{1}{\Delta^2}, \qquad \text{with} \quad \Delta = \min_\mathbf{k}\left|E_{m,\mathbf{k}} - E_{n,\mathbf{k}}\right| \quad \text{and } m\mathbf{k} \neq n\mathbf{k}. \quad \text{(B7)}$$

Combining Eqs. (B6)-(B7) and substituting into Eq. (B3) one arrives at the desired bound on the magnitude of the Berry curvature dipole:

$$\left|D_{ij}\right| \le \frac{S_F}{4\pi^3}\frac{\hbar^2 v_{\max}^2}{\Delta^2}. \tag{B8}$$

Here, $v_{\max} = \max_r \left\|\hat{v}_{\mathbf{k},r}\right\|_\infty$ can be understood as the maximum velocity of the Bloch electrons. Note that $v_{\max}$ is finite in typical tight-binding models, and usually on the order of $10^4 - 10^6\, m/s$. In the bound (B8) the gap width can be defined locally with respect to the Fermi surface, such that $\Delta = \min_{E_{n\mathbf{k}}=\mu}\left|E_{m,\mathbf{k}} - \mu\right|$ and $m\mathbf{k} \neq n\mathbf{k}$.

Equation (B8) provides a particularly simple bound for the Berry curvature dipole in terms of the area of the Fermi surface, the maximum velocity of the Bloch electrons, and the minimum energy separation from neighboring bands. The scale $\frac{1}{\Delta^2}$ is consistent with known analytical models; see for example Ref. [61] for BiTeI.

## Appendix C: Floquet expansion

In this Appendix, I obtain a formal solution for the system (13) by expanding the unknown fields $\mathbf{P}_{\text{c,probe}}^{\text{int}}$ and $\tilde{\mathbf{E}}_{\text{probe}}$ in a Floquet series:

$$\tilde{\mathbf{E}}_{\text{probe}} = \sum_n \tilde{\mathbf{E}}_n e^{-i(\omega+n\Omega)t}, \qquad \mathbf{P}_{\text{c,probe}}^{\text{int}} = \sum_n \mathbf{P}_n e^{-i(\omega+n\Omega)t}. \tag{C1}$$

Plugging these expansions and Eqs. (8) and (9) into Eq. (13), assuming a probe field of the form $\mathbf{E}_{\text{probe}}(t) = \mathbf{E}_{0\omega} e^{-i\omega t}$, it is found that:

$$-i\omega_n \frac{\mathbf{P}_n}{\varepsilon_0} = \frac{\varepsilon_\infty \omega_{\mathrm{ENZ}}^2}{\Gamma} \tilde{\mathbf{E}}_n - \frac{e^3}{2\varepsilon_0 \hbar^2 \Gamma} \left( \tilde{\mathbf{E}}_{n-1} \cdot \overline{\mathbf{D}} \right) \times \mathbf{E}_{\mathrm{pu}0}^{\mathrm{int}} - \frac{e^3}{2\varepsilon_0 \hbar^2 \Gamma} \left( \tilde{\mathbf{E}}_{n+1} \cdot \overline{\mathbf{D}} \right) \times \mathbf{E}_{\mathrm{pu}0}^{\mathrm{int}*}$$

$$- \frac{e^3}{2\varepsilon_0 \hbar^2 \Gamma} \left( \tilde{\mathbf{E}}_{\mathrm{pu}0} \cdot \overline{\mathbf{D}} \right) \times \tilde{\mathbf{L}} \cdot \left( \mathbf{E}_{0\omega} \delta_{n-1,0} - \mathbf{L} \cdot \frac{\mathbf{P}_{n-1}}{\varepsilon_0} \right) \tag{C2a}$$

$$- \frac{e^3}{2\varepsilon_0 \hbar^2 \Gamma} \left( \tilde{\mathbf{E}}_{\mathrm{pu}0}^* \cdot \overline{\mathbf{D}} \right) \times \tilde{\mathbf{L}} \cdot \left( \mathbf{E}_{0\omega} \delta_{n+1,0} - \mathbf{L} \cdot \frac{\mathbf{P}_{n+1}}{\varepsilon_0} \right).$$

$$\tilde{\mathbf{E}}_n = \frac{\Gamma}{\Gamma - i\omega_n} \tilde{\mathbf{L}} \cdot \left( \mathbf{E}_{0\omega} \delta_{n0} - \mathbf{L} \cdot \frac{\mathbf{P}_n}{\varepsilon_0} \right). \tag{C2b}$$

where $\omega_n = \omega + n\Omega$, $n = 0, \pm 1, \pm 2, ...$ is a generic integer, and $\tilde{\mathbf{L}} = \left( \mathbf{1} + (\varepsilon_\infty - 1) \mathbf{L} \right)^{-1}$. By eliminating $\tilde{\mathbf{E}}_n$, one obtains an infinite-dimensional linear system for $\mathbf{P}_n$:

$$\left( -i\omega_n \mathbf{1} + \frac{\varepsilon_\infty \omega_{\mathrm{ENZ}}^2}{\Gamma - i\omega_n} \tilde{\mathbf{L}} \cdot \mathbf{L} \right) \cdot \mathbf{P}_n$$

$$- \frac{e^3}{2\varepsilon_0 \hbar^2 \Gamma} \left[ \left( \tilde{\mathbf{E}}_{\mathrm{pu}0} \cdot \overline{\mathbf{D}} \right) \times \tilde{\mathbf{L}} \cdot \mathbf{L} \cdot \mathbf{P}_{n-1} + \left( \tilde{\mathbf{E}}_{\mathrm{pu}0}^* \cdot \overline{\mathbf{D}} \right) \times \tilde{\mathbf{L}} \cdot \mathbf{L} \cdot \mathbf{P}_{n+1} \right] \tag{C3a}$$

$$+ \frac{e^3}{2\varepsilon_0 \hbar^2 \Gamma} \left[ \frac{\Gamma}{\Gamma - i\omega_{n-1}} \mathbf{E}_{\mathrm{pu}0}^{\mathrm{int}} \times \left( \overline{\mathbf{D}}^{\mathrm{T}} \cdot \tilde{\mathbf{L}} \cdot \mathbf{L} \cdot \mathbf{P}_{n-1} \right) + \frac{\Gamma}{\Gamma - i\omega_{n+1}} \mathbf{E}_{\mathrm{pu}0}^{\mathrm{int}*} \times \left( \overline{\mathbf{D}}^{\mathrm{T}} \cdot \tilde{\mathbf{L}} \cdot \mathbf{L} \cdot \mathbf{P}_{n+1} \right) \right] = \varepsilon_0 \mathbf{Q}_n$$

with the independent term given by:

$$\mathbf{Q}_n = \frac{\varepsilon_\infty \omega_{\mathrm{ENZ}}^2}{\Gamma - i\omega} \tilde{\mathbf{L}} \cdot \mathbf{E}_{0\omega} \delta_{n0}$$

$$- \frac{e^3}{2\varepsilon_0 \hbar^2 \Gamma} \left[ \delta_{n-1,0} \left( \tilde{\mathbf{E}}_{\mathrm{pu}0} \cdot \overline{\mathbf{D}} \right) \times \tilde{\mathbf{L}} \cdot \mathbf{E}_{0\omega} + \delta_{n+1,0} \left( \tilde{\mathbf{E}}_{\mathrm{pu}0}^* \cdot \overline{\mathbf{D}} \right) \times \tilde{\mathbf{L}} \cdot \mathbf{E}_{0\omega} \right] \tag{C3b}$$

$$- \frac{e^3}{2\varepsilon_0 \hbar^2 \Gamma} \left[ \delta_{n-1,0} \frac{\Gamma}{\Gamma - i\omega} \left( \left( \tilde{\mathbf{L}} \cdot \mathbf{E}_{0\omega} \right) \cdot \overline{\mathbf{D}} \right) \times \mathbf{E}_{\mathrm{pu}0}^{\mathrm{int}} + \delta_{n+1,0} \frac{\Gamma}{\Gamma - i\omega} \left( \left( \tilde{\mathbf{L}} \cdot \mathbf{E}_{0\omega} \right) \cdot \overline{\mathbf{D}} \right) \times \mathbf{E}_{\mathrm{pu}0}^{\mathrm{int}*} \right].$$

In practice, the above system is truncated and solved numerically. In all the numerical results of the article, only the $|n| \leq 2$ harmonics were retained. Including additional harmonics produced negligible changes.